\documentclass[10.5pt,letterpaper]{article}
\pdfoutput=1 

\usepackage[utf8]{inputenc} 

\usepackage[top=2.5cm, bottom=3cm, left=3.5cm, right=3.5cm,
           heightrounded,marginparwidth=1.5cm, marginparsep=1cm]{geometry} 

\usepackage{changepage} 
\usepackage{etoolbox}

\usepackage[shortlabels]{enumitem} 

\usepackage[square,numbers,merge,comma,sort&compress]{natbib} 
\makeatletter
\def\NAT@spacechar{\,}  
\makeatother

\usepackage{amsmath,amssymb,amsfonts,amsthm}
\usepackage{mathtools} 
\usepackage{old-arrows}
\usepackage[scale=0.7]{miama} 
\usepackage{slashed,cancel}
\usepackage{comment}   
\usepackage{relsize}   
\usepackage{setspace}  
\usepackage{moresize}  
\usepackage{epsfig}
\usepackage{latexsym}
\usepackage{mathrsfs,calligra,aurical} 
\usepackage{calc}
\usepackage{float}
\usepackage{appendix}
\usepackage{xargs}
\usepackage{extarrows}
\usepackage{empheq}
\usepackage{scalerel}

\usepackage{graphicx} 

\usepackage[labelsep=colon]{caption}  
\usepackage[labelsep=colon,aboveskip=10pt,belowskip=10pt]{subcaption}
\usepackage{array,multirow,makecell,booktabs}  
\newcolumntype{X}[2]{>{\centering\arraybackslash$}#1{#2\linewidth}<{$}}
\newcolumntype{R}[1]{>{\raggedleft\arraybackslash$}m{#1\linewidth}<{$}}
\newcolumntype{L}[1]{>{\raggedright\arraybackslash}m{#1\linewidth}}

\usepackage{titlesec}

\titleformat{\section}{\normalfont\large\bfseries}{\thesection}{1em}{}
\titleformat{\subsection}{\normalfont\normalsize\bfseries}{\thesubsection}{0.75em}{}
\titleformat{\subsubsection}{\normalfont\normalsize\bfseries}{\thesubsubsection}{0.75em}{}

\titlespacing*{\section}{0pt}%
                {4ex plus 1ex minus .5ex}{1.75ex plus .25ex minus .25ex}
\titlespacing*{\subsection}{0pt}%
                {3.5ex plus 1ex minus .5ex}{1.25ex plus .2ex minus .2ex}
\titlespacing*{\subsubsection}{0pt}%
                {2.5ex plus 0.75ex minus .2ex}{0.75ex plus .15ex minus .15ex}
\titlespacing*{\paragraph}{0pt}%
                {1.85ex plus 0.5ex minus .15ex}{1em}

\usepackage{titletoc}

\addtocontents{toc}{\addvspace{-0.75em}}  

\titlecontents{section}
  [1.25em] {\addvspace{0.7em plus 0pt}\small}
  {\thecontentslabel\hspace{0.75em}}{}
  {\hspace{0.5em}\titlerule*[0.5em]{.}\contentspage}
  [\addvspace{0.0em plus 0pt}]

\titlecontents{subsection}
  [2.75em] {\addvspace{0.075em plus0pt}\fns}
  {\thecontentslabel\hspace{0.75em}}{\thecontentslabel\hspace{0.75em}}
  {\hspace{0.5em}\titlerule*[0.5em]{.}\small\contentspage}
  [\addvspace{0.075em plus 0pt}]

\usepackage{environ}
\makeatletter
\NewEnviron{subalign}[1]{
\begin{subequations}\label{#1}
\begin{align} \BODY \end{align}
\end{subequations}      }
\makeatother
\makeatletter
{\begingroup%
\setlength{\abovedisplayskip}{10pt plus 4pt minus 9pt}%
\setlength{\abovedisplayshortskip}{0pt plus 2pt minus 2pt}%
\setlength{\belowdisplayskip}{12pt plus 3pt minus 9pt}%
\setlength{\belowdisplayshortskip}{7pt plus 3pt minus 4pt}%
\begin{subequations}%
}%
{\end{subequations}\ignorespacesafterend%
\endgroup}%
\makeatother

\makeatletter
{\begingroup%
\setlength{\abovedisplayskip}{{#1}pt plus 2pt minus 9pt}%
\setlength{\abovedisplayshortskip}{0pt plus 0pt minus 2pt}%
\setlength{\belowdisplayskip}{{#2}pt plus 3pt minus 9pt}%
\setlength{\belowdisplayshortskip}{7pt plus 3pt minus 4pt}%
\begin{subequations}%
}%
{\end{subequations}\ignorespacesafterend%
\endgroup}%
\makeatother

\makeatletter
{\begingroup%
\setlength{\abovedisplayskip}{{#1}pt plus 3pt minus 9pt}%
\setlength{\abovedisplayshortskip}{0pt plus 3pt}%
\setlength{\belowdisplayskip}{{#2}pt plus 3pt minus 9pt}%
\setlength{\belowdisplayshortskip}{7pt plus 3pt minus 4pt}%
\begin{equation}%
}%
{\end{equation}\ignorespacesafterend%
\endgroup}%
\makeatother

\makeatletter
{%
\begin{equation}%
\begin{split}{#1}%
}%
{\end{split}%
\end{equation}\ignorespacesafterend%
}%
\makeatother

\usepackage[svgnames]{xcolor}
\definecolor{Green}{rgb}{0.05, 0.45, 0.25}
\definecolor{dogwoodrose}{rgb}{0.8, 0.1, 0.55}
\definecolor{RRed}{rgb}{0.7, 0.1, 0.525}

\usepackage{bm}  
\usepackage{dsfont}  

\DeclareMathAlphabet{\mathpzc}{OT1}{pzc}{m}{it}
\DeclareMathAlphabet{\mathcal}{OMS}{cmsy}{m}{n}
\DeclareSymbolFontAlphabet{\Scr}{rsfs}
\DeclareMathAlphabet{\mathbold}{U}{BOONDOX-ds}{m}{n}
\SetMathAlphabet{\mathbold}{bold}{U}{BOONDOX-ds}{b}{n}
\DeclareMathAlphabet{\mathcalboondox}{U}{BOONDOX-calo}{m}{n}
\SetMathAlphabet{\mathcalboondox}{bold}{U}{BOONDOX-calo}{b}{n}
\DeclareMathAlphabet{\mathbcalboondox}{U}{BOONDOX-calo}{b}{n}

\newcommand\eqlinkcol{RRed}



\usepackage[breaklinks=true,backref=page]{hyperref}
\hypersetup{
    bookmarks=true,         
    pdfmenubar=true,        
    pdffitwindow=false,     
    pdfpagemode={UseNone},
    pdfstartview={FitH},
    pdfauthor={},     
    pdfsubject={},   
    pdfcreator={},   
    pdfproducer={},  
    colorlinks=true,
    bookmarks=true,
    bookmarksnumbered=true,
    plainpages,
    a4paper,
    linktoc=page,
    citecolor=blue,
    filecolor=black,
    linkcolor=\eqlinkcol,
    urlcolor=Green,
}
\renewcommand*{\backref}[1]{}
\renewcommand*{\backrefalt}[4]{%
\ifcase #1 %
\relax
\or
~{\small [\textsc{p.~\fns{\!#2}}]}
\else
~{\small [\textsc{p.~\fns{\!#2}}]}%
\fi}

\usepackage{footnotebackref}
\usepackage{hypernat} 

\makeatletter
\normalsize
\makeatother

\def\+{\:+\:}
\def\-{\:-\:}
\def\={\:=\:}
\def\'{``}
\def\*{{}^*}
\newcommand\fns{\footnotesize}
\newcommand\qqquad{\quad\quad\quad}

\providecommand{\abs}[1]{\lvert#1\rvert}

\newcommand\eps{\epsilon}
\newcommand\veps{\varepsilon}
\newcommand\w{\omega}

\newcommand\N{\mathcal{N}}
\newcommand\eD{e_{\textsc{d}}}

\newcommand\D{\mathcal{D}}

\newcommand\numinus{\nu_{{\!}_{-}}}
\newcommand\nuplus{\nu_{{\!}_{+}}}
\newcommand\rplus{r_{{\!}_{+}}}

\newcommand\M{\mathcal{M}}

\newcommand\SO{\mathrm{SO}}

\newcommand\V{\mathcal{V}}

\newcommand\X{\mathcal{X}}
\newcommand\Ucal{\mathcal{U}}
\newcommand\zb{{\bar{z}}}

\newcommandx{\sh}[1][1=\alpha,usedefault]{\sinh\left(#1\right)}
\newcommandx{\ch}[1][1=\alpha,usedefault]{\cosh\left(#1\right)}
\newcommandx{\sech}[1][1=\alpha,usedefault]{\operatorname{sech}\left(#1\right)}
\newcommandx{\cosech}[1][1=\alpha,usedefault]{\operatorname{cosech}\left(#1\right)}
\newcommandx{\tts}[1]{\text{\textsmaller{#1}}}
\newcommandx{\dm}[1][1=\mu,usedefault]{\partial_{#1}}
\newcommandx{\dmup}[1][1=\mu,usedefault]{\partial^{#1}}
\newcommandx{\subm}[2][1=p,2=A,usedefault]{{#1}_{\!\mathsmaller{#2}}}
\newcommandx{\subt}[2][1=p,2=A,usedefault]{{#1}_\text{\textsmaller{#2}}}
\newcommandx{\supm}[2][1=p,2=A,usedefault]{{#1}^{\!\mathsmaller{#2}}}
\newcommandx{\supt}[2][1=p,2=A,usedefault]{{#1}^\text{\textsmaller{#2}}}
\newcommandx{\subpt}[3][1=p,2=A,3=B,usedefault]{{#1}^\text{\textsmaller{#3}}_\text{\textsmaller{#2}}}
\newcommandx{\subpm}[3][1=p,2=A,3=B,usedefault]{{#1}^{\mathsmaller{#3}}_{\mathsmaller{#2}}}
\newcommandx{\LCTd}[4][1=\mu,2=\nu,3=\rho,4=\sigma,usedefault]{\veps_{#1#2#3#4}}
\newcommandx{\LCTu}[4][1=\mu,2=\nu,3=\rho,4=\sigma,usedefault]{\veps^{#1#2#3#4}}
\newcommandx{\gmetr}[2][1=\mu,2=\nu,usedefault]{g_{{#1}{#2}}}
\newcommandx{\invgmetr}[2][1=\mu,2=\nu,usedefault]{g^{{#1}{#2}}}
\newcommandx{\spc}[3][1=\mu,2=a,3=b,usedefault]{{\w_{#1}}^{\!\!{#2}{#3}}}
\newcommandx{\Conn}[3][1=\mu,2=\nu,3=\lambda,usedefault]{{\Gamma_{{#1}{#2}}}^{\!\!#3}}
\newcommandx{\viel}[2][1=\mu,2=a,usedefault]{{e_{#1}}^{\!#2}}
\newcommandx{\inviel}[2][1=a,2=\mu,usedefault]{{e_{#1}}^{#2}}
\newcommandx{\vieluu}[2][1=\mu,2=a,usedefault]{e^{#1#2}}
\newcommandx{\Rdduu}[4][1=\mu,2=\nu,3=a,4=b,usedefault]{{R_{{#1}{#2}}}^{{#3}{#4}}}

\newcommandx{\overbar}[1]{\mkern                   1.5mu\overline{\mkern-2.0mu#1\mkern-2.0mu}\mkern 1.5mu}
\newcommandx{\overbarcal}[1]{\mkern                   6.0mu\overline{\mkern-5.5mu#1\mkern-1.0mu}\mkern 1.5mu}  

\DeclareFixedFont\trfont{OT1}{phv}{b}{sc}{11}

\title{%
       \centering\boldmath\LARGE\bfseries%
       Instability of
       supersymmetric black holes via quantum phase transitions
       \vspace{1em}
       }

\usepackage[blocks]{authblk}  

\author[a,c]{Andrés Anabalón%
    \thanks{{\href{mailto:andres.anabalon@uai.cl}{\texttt{andres.anabalon@uai.cl}}}}%
    }

\author[b]{Dumitru Astefanesei%
    \thanks{{\href{mailto:dumitru.astefanesei@pucv.cl}{\texttt{dumitru.astefanesei@pucv.cl}}}}%
    }

\author[c,d]{Antonio Gallerati%
    \thanks{{\href{mailto:antonio.gallerati@polito.it}{\texttt{antonio.gallerati@polito.it}}}}%
    }

\author[c,d]{Mario Trigiante%
    \thanks{{\href{mailto:mario.trigiante@polito.it}{\texttt{mario.trigiante@polito.it}}}}%
    }

\affil[a]{%
{Universidad Adolfo Ibáñez, Dep.\ de Ciencias, Fac.\ de Artes Liberales. Av.\ Padre Hurtado 750, Viña del Mar, Chile}\smallskip}

\affil[b]{%
{Pontificia Universidad Católica de Valparaíso, Instituto de Fısica - Valparaíso, Chile.}\smallskip}

\affil[c]{Politecnico di Torino, Dipartimento DISAT. Corso Duca degli Abruzzi 24, 10129 Torino, Italy\smallskip}

\affil[d]{Istituto Nazionale di Fisica Nucleare (INFN), sez.\ TO. Via Pietro Giuria 1, Torino, Italy}

\date{}

\makeatletter
\patchcmd{\@maketitle}{\begin{center}}{\begin{adjustwidth}{-0.25in}{-0.25in}\begin{center}}{}{}
\patchcmd{\@maketitle}{\end{center}}{\end{center}\end{adjustwidth}}{}{}
\makeatother

\begin{document}

\maketitle


\begin{abstract}
{\noindent
In this paper we prove that the four-dimensional hyperbolic supersymmetric black holes can be unstable in the canonical ensemble. To this end, we work with an infinite class of \,$\N=2$ supergravity theories interpolating between all the single dilaton truncations of the $\SO(8)$ gauged \,$\N=8$ supergravity. Within these models, we study electrically charged solutions of two different kinds: supersymmetric hairy and extremal non-supersymmetric Reissner-Nordstr\"{o}m black holes. We consider these solutions within the same canonical ensemble and show that, for suitable choices of the parameters defining the \,$\N=2$ model, the supersymmetric solution features a higher free energy than the non-supersymmetric one. In the absence of additional selection rules, this would imply an instability of the supersymmetric configuration, hinting towards a possible supersymmetry breaking mechanism.
}
\end{abstract}


\tableofcontents

\pagebreak


\section{Introduction}
Supersymmetry has been playing a special role in the study of quantum properties of gravity, mainly because of its beneficial effects in taming ultraviolet divergences in the quantum theory. Related to this, the existence of an amount of supersymmetry preserved by a supergravity vacuum solution was shown to imply the stability of the latter, by virtue of positive-energy theorems \cite{Witten:1981mf, Breitenlohner:1982jf, Gibbons:1983aq}.\par
An open question arises as to whether, in a theory with a maximally-supersymmetric vacuum, supersymmetry of a non-vacuum solution can prevent it from decaying into a non-supersymmetric one. A necessary condition for this to occur is that the free-energy of the former exceeds that of the latter. Remarkably, in this letter, we find that, for a specific class of supergravity theories \cite{Anabalon:2020pez}, a supersymmetric hairy black hole whose thermodynamic free energy exceeds that of a non-supersymmetric Reissner-Nordstr\"{o}m-AdS extremal solution within the same canonical ensemble, defined by fixed common values of their charges.\par
Phase transitions between different gravitational solutions are well known to take place and date back to the article of Hawking and Page \cite{Hawking:1982dh}. Theories with scalars are particularly suited to provide a setup where second order phase transitions can occur. This happens because the scalar field does not provide an extra conserved charge, while, at the same time, it modifies the space of solutions in such a way as to allow a hairy black hole configuration \cite{Martinez:2004nb,Hartnoll:2008vx,Anabalon:2012sn,Gallerati:2019mzs}. Hence, for the same values of the charges, there are at least two solutions, the hairy and the bald black hole. A second order phase transition can take place if these phases can coexist, and a first order phase transition would happen whenever they feature different free energies. The main objective of this paper is to apply these concepts to the stability of supersymmetric solutions. We would like to remark that the coexistence of supersymmetric phases was recently spotted in the absence of scalar fields in \cite{Anabalon:2021tua}.\par
The class of models that will be considered here belongs to four-dimensional, $\N=2$ gauged supergravity coupled to one vector multiplet in the presence of an abelian Fayet-Iliopoulos dyonic gauging. These models differ in the geometry of the scalar manifold, whose prepotential depends on the complex scalar through a power-law defined by a parameter $\nu$ \cite{Anabalon:2017yhv,Anabalon:2020pez}%
\footnote{%
for certain values of $\nu$, single-scalar truncations of these models are also truncations \cite{Anabalon:2020pez,Gallerati:2021cty} of a gauged maximal supergravity \cite{deWit:1981sst,deWit:1982bul,DallAgata:2012mfj,
Trigiante:2016mnt,Gallerati:2016oyo}.
}.
Static hairy single-scalar black holes within these models were recently constructed and studied in \cite{Anabalon:2020pez,Gallerati:2021cty}. In the BPS limit \cite{Anabalon:2020pez}, such solutions are related to the class of black holes considered in \cite{Cacciatori:2009iz,Hristov:2010eu,Hristov:2010ri,Hristov:2011ye,
Toldo:2012ec,Chow:2013gba,Gnecchi:2013mja,Gnecchi:2013mta,Lu:2014fpa, Chimento:2015rra,Klemm:2015xda,Faedo:2015jqa,Hristov:2018spe, Daniele:2019rpr,Anabalon:2020qux}.
Hairy black hole solutions in higher dimensions were considered in \cite{Feng:2013tza}, generalizing those found in \cite{Anabalon:2012ta}.\par
In this work, we consider the electrically-charged, single-scalar BPS black holes of \cite{Anabalon:2020pez}, as well as extremal Reissner-Nordstr\"{o}m-AdS black holes \cite{Romans:1991nq}, characterized by the same values of the charges.
Therefore, our set-up can support both supersymmetric and SUSY broken phases and so it is suitable for investigating possible quantum phase transitions. We show that for certain values of the parameters defining the model under consideration, for hyperbolic horizons the free-energy of the hairy solution exceeds the one of the (non-supersymmetric) extremal Reissner-Nordstr\"{o}m-AdS. This hints towards an instability of the former supersymmetric solution, unless some kind of selection rule is at work to prevent its decay.
We will also briefly discuss the existence of asymptotic Killing spinors, which could define an underlying supersymmetry algebra, depending on the global geometry of the horizon of the solution. This would be instrumental for the discussion of the thermodynamic stability of the backgrounds.

\section{The model}
We are going to consider the simplest supersymmetric model with electromagnetic fields that would allow to have more than one black hole solution for the same boundary conditions. To this end, we need one scalar field and, therefore, supersymmetry requires the introduction of two gauge fields. 
Furthermore, in order to make contact with M-theory, we need the model to be a subsector of the maximal supergravity in four dimensions.\par
A minimal model that satisfies the above requirements was found in \cite{Anabalon:2020pez}. The framework under consideration is an $\mathcal{N}=2$ supergravity theory with no hypermultiplets and a single vector multiplet containing a complex scalar field $z$. The geometry of the special K\"ahler manifold is characterized by a prepotential of the form:
\begin{equation}
\mathcal{F}(\X^\Lambda)\:=-\frac{i}{4}\:\left(\X^0\right)^{n}\left(\X^1\right)^{2-n}\,,
\label{eq:prepotentialn}
\end{equation}
$\X^\Lambda(z)$ being components of a holomorphic section of the symplectic bundle over the manifold. The coordinate $z$ is identified with the ratio $\X^1/\X^0$, in a local patch in which $\X^0\neq 0$.
If we set $\X^0=1$, the holomorphic section $\Omega^M$ and the K\"ahler potential of the model read:
\begin{equation}
\Omega^M=
\left(\begin{matrix}
1  \cr  z  \cr -\dfrac{i}{4}\,n\,z^{2-n} \cr -\dfrac{i}{4}\,(2-n)\,z^{1-n}
\end{matrix}\right)\,,
\qqquad
e^{-\mathcal{K}}\=\frac{1}{4}\,z^{1-n}\,\big(n\,z-(n-2)\,\bar{z}\big)\+\text{c.c.}
\end{equation}
The theory is deformed by the introduction of abelian electric-magnetic FI terms defined by a constant symplectic vector $\theta_M$\,,
\begin{equation}
\theta_M=\left(\theta_1,\,\theta_2,\,\theta_3,\,\theta_4\right)\,, \end{equation}
encoding the gauge parameters of the model. The scalar potential $V(z,\zb)$ can be then obtained from:
\begin{equation}
V\=\left(g^{i\bar{\jmath}}\,\Ucal_i^M\,\overbar{\Ucal}_{\bar{\jmath}}^N
         -3\,\V^M\,\overbar{\V}^N\right)\theta_M\,\theta_N\,=\,
    -\frac{1}{2}\,\theta_M\,\M^{MN}\,\theta_N-4\,\V^M\,\overbar{\V}^N\theta_M\,\theta_N\;,
\label{eq:Vpot}
\end{equation}
where \:$\V^M=e^{\mathcal{K}/2}\,\Omega^M$, \;$\Ucal_i^M=\D_i\,\V^M$\, and \,$\M(\phi)$\, is the symplectic, symmetric, negative definite matrix encoding the non-minimal couplings of the scalars to the vector fields of the theory.\par\smallskip
The complex scalar can be expressed as \,$z=e^{\lambda\,\phi}+i\,\chi$\,, with $\lambda=\sqrt{\tfrac{2}{n(2-n)}}$\,.\: In particular, we restrict to a truncation to the dilaton field $\phi$ only $(\chi=0)$.%
\footnote{%
It is possible to explicitly verify that the truncation to the dilaton field is consistent by inspecting the consistency of the axion field equation after the $\chi=0$ truncation \cite{Anabalon:2020pez}.
}\:
We then make the shift
\begin{equation}
\phi\;\rightarrow\;\phi-\frac{2\,\nu}{\lambda\,(1+\nu)}\,\log(\theta_2\,\xi)\,,
\label{eq:phishift}
\end{equation}
that gives a minimum of the potential for $\phi=0$ (see also subsequent \eqref{eq:Vphi} and \eqref{eq:dVphi}) and redefine the FI terms as:
\begin{equation} \label{eq:newpar}
\theta_{1}=\frac{1+\nu}{-1+\nu}\;\theta_{2}^{-\frac{-1+\nu}{1+\nu}}\,\xi^{-\frac{2\,\nu}{1+\nu}}\,,\qquad
\theta_{3}=2\,\alpha\left(\xi\,\theta_{2}\right)^{\frac{-1+\nu}{1+\nu}}\,,\qquad
\theta_{4}=\frac{2\,\alpha}{\theta_{2}\,\xi}\,,
\end{equation}
having introduced the quantities
\begin{equation}
\nu=\frac{1}{-1+n}\:,\qqquad
\xi\=\frac{2\,L\,\nu}{-1+\nu}\,\frac{1}{\sqrt{1-\alpha^{2}\,L^{2}}}\:,
\end{equation}
$L$ being the AdS radius and the parameter $\alpha$ controlling the strength of the dyonic gauging.\par\smallskip
After the shift \eqref{eq:phishift}, the scalar field $z$ is expressed as
\begin{equation}
z\=\left(\theta_{2}\,\xi\right)^{-\frac{2\,\nu}{1+\nu}}\,e^{\lambda\,\phi}\,, \end{equation}
and the potential \eqref{eq:Vpot} explicitly reads \cite{Anabalon:2020pez}
\begingroup
\belowdisplayskip=-5pt
\belowdisplayshortskip=-5pt
\begin{equation}
\begin{split}
V(\phi)\:=\,&-\frac{\alpha^2}{\nu^2}\,\left(\frac{(-1+\nu)(-2+\nu)}{2}\,
    e^{-\phi\,\ell\,(1+\nu)} + 2\,(-1+\nu^2)\,e^{-\phi\,\ell} +\frac{(1+\nu)(2+\nu)}{2}\,e^{\phi\,\ell\,(-1+\nu)}\right)+
\\
    &+\frac{\alpha^2-L^{-2}}{\nu^2}\,\left(\frac{(-1+\nu)(-2+\nu)}{2}\,
    e^{\phi\,\ell\,(1+\nu)} + 2\,(-1+\nu^2)\,e^{\phi\,\ell} +\frac{(1+\nu)(2+\nu)}{2}\,e^{-\phi\,\ell\,(-1+\nu)}\right),
\end{split}
\label{eq:Vphi}
\end{equation}
\endgroup
where \,$\ell=\tfrac{\lambda}{\nu}=\tfrac{1}{\nu}\,\sqrt{\tfrac{2\,\nu^2}{-1+\nu^2}}$\, and having disposed of $\theta_2$ by the above redefinitions.\par\medskip
After the restriction to the dilaton truncation, the action takes the form
\begin{equation}
\mathscr{S}=\frac{1}{8\pi G}\,\int{\!
    d^4x\;\eD\left(-\,\frac{R}{2}\,+\,\frac{\partial_{\mu}\phi\,\partial^{\mu}\phi}{2}
    \,-\,\frac{1}{4}\,e^{(-1+\nu)\,\ell\,\phi}\,\left(F^{1}\right)^{2}
    -\frac{1}{4}\,e^{-(1+\nu)\,\ell\,\phi}\,\left(F^{2}\right)^{2}
    -\,V(\phi)\right)}\,,
\label{eq:action}
\end{equation}
in terms of canonically normalized field strengths expressed as
\begin{equation}
F_{\mu\nu}^{\Lambda}=\partial_{\mu}A_{\nu}^{\Lambda}-\partial_{\nu}A_{\mu}^{\Lambda}\:.
\end{equation}
The $\nu$ parameter, $\abs{\nu}>1$, is a real parameter that interpolates between all the single dilaton truncations of the maximal $\SO(8)$ supergravity in four dimensions. These truncations break $\SO(8)$ as follows \cite{Anabalon:2020pez,Gallerati:2021cty}:
\begin{equation}\label{eq:SO8breaking}
\begin{alignedat}{3}
\nu=\frac{4}{3} & \;\;\rightarrow\quad && \SO(7) \,,             \\[0.5ex]
\nu=2           & \;\;\rightarrow\quad && \SO(6)\times\SO(2) \,, \\[1ex]
\nu=4           & \;\;\rightarrow\quad && \SO(5)\times\SO(3) \,, \\[1ex]
\nu=\infty      & \;\;\rightarrow\quad && \SO(4)\times\SO(4) \,,
\end{alignedat}
\end{equation}
while all the other values of $\abs{\nu}>1$ are $\mathcal{N}=2$ supergravity
theories.\par\smallskip
The scalar field potential satisfies
\begin{equation}
V(0)=-\frac{3}{L^{2}}\,,\qquad\qquad
\left.\frac{dV(\phi)}{d\phi}\right\vert_{\phi=0}\!=0\,,\qquad\qquad
\left.\frac{d^{2}V(\phi)}{d\phi^{2}}\right\vert_{\phi=0}\!=-\frac{2}{L^{2}}\,.
\label{eq:dVphi}
\end{equation}
and the equation for the dilaton is
\begin{equation}
-\Box\phi+\frac{dV(\phi)}{d\phi}-\,\frac{(-1+\nu)\,\ell}{4}
\:e^{(-1+\nu)\,\ell\,\phi}\,\left(F^{1}\right)^{2}
+\frac{(1+\nu)\,\ell}{4}\:e^{-(1+\nu)\,\ell\,\phi}\,\left(F^{2}\right)^{2}\=0\,.
\end{equation}
Therefore, when $\phi=0$, this equation is satisfied if
\begin{equation}
F_{\mu\nu}^{1}\=\pm\,\sqrt{\frac{1+\nu}{-1+\nu}}\,F_{\mu\nu}^{2}\:,
\label{eq:constr}
\end{equation}
which implies that there is a single gauge field when $\phi=0$. Hence, the
Lagrangian \eqref{eq:action} has at least two solutions for a given set of boundary conditions which are compatible with $\phi=0$.\par

\paragraph{Hairy vs.\ Reissner-Nordstr\"{o}m.}
Let us consider the case when the system is fully characterized by its electric charge and the total energy. In this case, when $\phi=0$, the field equations are satisfied by the well known Reissner-Nordstr\"{o}m solution, while when $\phi\neq0$ one obtains the hairy black holes solutions of \cite{Anabalon:2020pez}. In the grand canonical ensemble, when the temperature of the system vanishes, there exist two black holes for the same chemical potential $\Phi$. On the other side, in the canonical ensemble, there exist two black holes for the same electric charge $Q$. Furthermore, the purely electrically charged, extremal Reissner-Nordstr\"{o}m black hole is not supersymmetric \cite{Romans:1991nq}.\par
However, the black holes with $\phi\neq0$ can be supersymmetric \cite{Anabalon:2020pez,Gallerati:2021cty}. Then, the study of the phase transition between these states becomes a way to study under which conditions supersymmetry is actually expected to be broken or unbroken. In the following sections, we mathematically describe these two states.

\section{The extremal Reissner-Nordstr\"{o}m black hole in AdS}
This solution is extremely well-known, for a discussion within the context
of the spherically symmetric case see \cite{Chamblin:1999tk,Chamblin:1999hg}. In general, the spacetime is foliated by \:$d\Sigma_k^{2}=d\theta^2+\tfrac{\sin^2(\sqrt{k}\,\theta)}{k}\,d\varphi^2$\:, namely the metric on the $2D$-surfaces $\Sigma_{k}=\{\mathbb{S}^{2},\,\mathbb{H}^{2},\,\mathbb{R}^{2}\}$
(sphere, hyperboloid and flat space) with constant scalar curvature
\thinspace $R=2\,k$.%
\footnote{%
To compare with our previous paper, \cite{Anabalon:2014fla}, note that the scalar curvature of $\Sigma _k$ has a different normalization.
}\par
A Reissner-Nordstr\"{o}m black hole solution for the action \eqref{eq:action} is written as:
\begin{equation}
ds^2\=f(r)\,dt^2-\frac{1}{f(r)}\,dr^2-r^2\,d\Sigma_k^2\,,\qquad\qquad
f(r)=k-\frac{m}{r}+\frac{L^2\,q^2}{r^2}+\frac{r^2}{L^2}\,,\quad
\end{equation}
with gauge fields
\begin{equation}
A^1_\mu=\left(\frac{L\,q\,\nuplus}{r},\,0,\,0,\,0\right),
\qqquad
A^2_\mu=\left(\epsilon_1\,\frac{L\,q\,\numinus}{r},\,0,\,0,\,0\right),
\end{equation}
and a convenient parametrization, describing an extremal configuration, is given by
\begin{equation}
m=\frac{2\,\rplus(k\,L^2+2\,\rplus^2)}{L^2}\,,
\qqquad
\nu_{{\!}_{\pm}}=\sqrt{1\pm\frac{1}{\nu}}\,,
\qqquad
\epsilon_{1}=\pm1\,.\qquad
\end{equation}
%
%
Constraint \eqref{eq:constr} implies that we have only one free integration constant $q$ associated to the electric charge of the solution. The latter can be suitably expressed as
\begin{equation}
q\=\frac{\rplus}{L^2}\:\sqrt{3\,\rplus^2+k\,L^2}
\;\;\quad\leftrightarrow\quad\;
\rplus\=L\:\sqrt{\frac{-k+\sqrt{k^2+12\,q^2}}{6}}\:,
\end{equation}
making manifest $\rplus$ as the horizon radius, once inserted the expression for $q$ in metric.
The physical mass and charges are given by
\begin{equation}
M=\frac{\sigma_{k}}{8\pi G}\:m\,,
\qqquad
q_1=\frac{\sigma_{k}}{8\pi G}\:L\:\nuplus\,q\,,
\qqquad
q_2=\frac{\sigma_{k}}{8\pi G}\:\eps_1\:L\:\numinus\,q\,,\qquad
\label{eq:Mq1q2}
\end{equation}
where $\sigma_{1}=4\pi$,\,  $\sigma_{-1}=8\pi(g-1)$ and $\sigma_{0}$ represent the volume of a two-dimensional plane.%
\footnote{%
A compact two-dimensional surface of genus $g$ is locally homeomorphic to a
surface of negative constant curvature with this volume.}
The entropy and the potentials of the extremal solution then reads
\begin{equation}\label{eq:SPhi1Phi2}
S=\frac{\sigma_{k}\;\rplus^2}{4\!\:G}\,,
\qquad\qquad
\Phi_1=\frac{L\,q}{\nuplus\,\rplus}\,,
\qquad\qquad
\Phi_2=\epsilon_{1}\:\frac{L\,q}{\numinus\,\rplus}\,,\qquad
\end{equation}
and the extremality condition for the first law can be directly checked as
\begin{equation}
\delta M\=\Phi_{1}\,\delta q_{1}\+\Phi_{2}\,\delta q_{2}\:,
\label{First Law}
\end{equation}
the charges $\left(q_{1},\,q_{2}\right)$ indeed satisfying constraint \eqref{eq:constr}.\par\medskip
We shall be interested in the canonical ensemble, where the relevant thermodynamic potential is the Helmholtz free energy:
\begin{equation}
F^\textsc{rn}\=E-T\,S\=E\:.
\end{equation}
For a field theory living on the three-dimensional surface
\begin{equation}
ds^{2}\=-dt^{2}+R^{2}\,d\Sigma_{k}\:,
\end{equation}
the field theory extensive thermodynamic variables are related as
\begin{equation}
X^\textsc{qft}=\tfrac{L}{R}\,X^\textsc{sugra}\:,
\end{equation}
and, for instance, the energy is
\begin{equation}
X^\textsc{qft}\=C\,N^{p}\,\frac{\sqrt{6}\,\sigma_{k}}{72\,\pi\,R}\:
    \frac{k\,\mathpzc{H}+12\,q^2}{\sqrt{\mathpzc{H}}}\:,
\end{equation}
the extensive thermodynamic variables being the same in the supergravity and
the field theory side. Note that we have used the holographic dictionary,
which relates gravitational and field theory quantities $\tfrac{L^{2}}{G}=C\,N^{p}$. Here $N$ is a large number, typically associated with the rank of a gauge group, $C$ is a number which depends on the theory and $p$ a positive number. For instance, it is known that $\tfrac{L^{2}}{G}=\tfrac{2\sqrt{2}}{3}\,K^{1/2}\,N^{3/2}$, with $K$ the level and $N$ the rank of the gauge groups of ABJM theory \cite{Aharony:2008ug,Aharony:2008gk}. We make this remark to emphasize that the supersymmetry breaking presented below is a mechanism that takes place in supergravity and equivalently, via AdS-CFT correspondence, in the dual quantum field theory.

\section{The hairy black holes}
Recently, we have constructed a family of electrically charged non-extremal black holes  in \cite{Anabalon:2020pez}.
Here, we express the electric family solution as%
\footnote{%
the field strengths $F_1$, $F_2$ are the canonically normalized quantities in the action \eqref{eq:action}, and were denoted by $\bar{F}_1$, $\bar{F}_2$ in \cite{Anabalon:2020pez}}:
\begin{subequations}\label{eq:elsol}
\begin{align}
&\phi\:=\,-\ell^{-1}\ln(x)\,,
\\[1.75ex]
&F^1_{tx}\,=\:Q_1\,x^{-1+\nu},
\qquad\;F^2_{tx}\,=\:Q_2\,x^{-1-\nu},
\\[2.5ex]
&f(x)\,=\:\frac{x^{2-\nu}\,\mu^{2}\,(-1+x^{\nu})^{2}}{\nu^{2}}\,k+\alpha^{2}L^{2}
    \,\left(-1+\frac{x^{2}}{\nu^{2}}\,\big(%
    \,(2+\nu)\,x^{-\nu}-(-2+\nu)\,x^{\nu}+\nu^{2}-4\,\big)\right)+
\nonumber\\
    &\phantom{f(x)\,=\:} +1+\frac{x^{2-\nu}\,\mu^{2}\,(-1+x^{\nu})^{3}}{\nu^{3}}
    \left(\frac{Q_{1}^{2}}{(1+\nu)}-\frac{Q_{2}^{2}}{(-1+\nu)}\,x^{-\nu}\right)\,,
\\[1.6\jot]
&\Upsilon (x)\,=\:\frac{x^{-1+\nu}\,\nu^{2}}{\mu^{2}\,(-1+x^{\nu})^{2}}\,,
\\[1.75ex]
&ds^{2}\,=\;\Upsilon(x)\left(f(x)\,dt^{2}-\frac{\mu^{2} L^2}{f(x)}\,dx^{2}
    -L^2\,d\Sigma_{k}\right)\,.  \label{subeq:metric}
\end{align}
\end{subequations}
in terms of the integration constant parameter $\mu$.

\paragraph{AdS boundary conditions, mass and thermodynamics.}
This analysis has been done in detail in \cite{Anabalon:2020pez}.
In particular, to make contact with the AdS canonical coordinates, we consider the following fall-off:
\begin{equation}
\Upsilon(x)\= \frac{r^2}{L^2} + O\left(r^{-2}\right)\;.
\end{equation}
The change of coordinates that provides the right asymptotic behaviour is
\begin{equation} \label{CC}
x\= 1 \pm
    \left(\frac{L}{\mu\,r}+L^3\,\frac{1-\nu^2}{24\,\left(\mu\,r\right)^3}\right)  +L^4\,\frac{\nu^2-1}{24\,\left(\mu\,r\right)^4}\;,
\end{equation}
where we take $\mu>0$ and the $\pm$ sign depends on whether one takes the $x<1$ ($-$) or $x>1$ ($+$). Accordingly, the asymptotic behaviour of the scalar field is
\begin{equation}\label{fall off}
\phi\=L^2\,\frac{\phi_0}{r}+L^4\,\frac{\phi_1}{r^2}+O\left(r^{-3}\right)\,=\,
    \mp L\,\frac{1}{\ell\,\mu\,r}+L^2\,\frac{1}{2\,\ell\,\mu^2\,r^2}+ O\left(r^{-3}\right)\;,
\end{equation}
where we have normalized $\phi_0$ and $\phi_1$ to match their conformal and engineering dimension. In the canonical coordinates, we can now easily read off the coefficients of the leading and subleading terms in the scalar boundary expansion
\begin{equation}
\phi_0=\mp\frac{1}{\ell\,\mu\,L}\;, \qquad
\phi_1=\frac{\ell}{2}\,\phi_{0}^2\;,
\label{eq:bc}
\end{equation}
which corresponds to AdS invariant boundary conditions \cite{Hertog:2004dr,Anabalon:2020pez}, namely a triple trace deformation in the boundary theory. Hence, the boundary conformal symmetry is preserved and the black hole mass can be extracted from the asymptotic expansion of the spacetime \cite{Henneaux:2006hk, Anabalon:2014fla,Anabalon:2015xvl,Anabalon:2017yhv,
Anabalon:2020pez,Anabalon:2020qux,Gallerati:2021cty}. The expansion of metric \eqref{eq:elsol} explicitly reads:
\begin{equation}
\begin{split}
g_{tt}&\=\frac{r^2}{L^2}+k-\frac{\mu_\textsc{e}\,L^4}{r}
    +O\left(r^{-2}\right)\;,\\[2\jot]
g_{rr}&\:=-\frac{L^2}{r^2}-L^6\:\frac{k\,L^{-2}+\frac{1}{2}\,\phi_0^2}{r^4}
          +O\left(r^{-5}\right)\;,
\end{split}
\end{equation}
where
\begin{equation}
\mu_\textsc{e}=\frac{\nu^2-4}{3\,\mu^3 L}\,\alpha^2-\frac{k}{\mu\,L^3}
    +\frac{Q_2^2}{(-1+\nu)\,\mu\,L^3}-\frac{Q_1^2}{(1+\nu)\,\mu\,L^3}\;,
\end{equation}
and the black hole mass then reads
\begin{equation}
M_\phi\=\frac{\sigma_k}{8\pi G}\:\mu_\textsc{e}\,L^4\:.
\end{equation}
The temperature is given by
\begin{equation}
T\=\frac{\abs{f(x)'}}{4\,\pi\,\mu\,L}\,\bigg|_{x=x_+}\;,
\end{equation}
where $f(x_+)=0$, while the entropy is expressed as
\begin{equation}
S_\phi\=\frac{\sigma_k}{8\pi G}\:2\pi\,L^2\,\Upsilon(x_+)\:.
\end{equation}
Finally, the charges and electric potentials are
\begin{equation}
\begin{split}
q_1^\phi&=\frac{\sigma_k}{8\pi G}\:\frac{L\,Q_1}{\mu}\,,
\qquad\;\;
\Phi_1^\phi=\frac{-1+x_{+}^\nu}{\nu}\:Q_1\,,
\\[2.5\jot]
q_2^\phi&=\frac{\sigma_k}{8\pi G}\:\frac{L\,Q_2}{\mu}\,,
\qquad\;\;
\Phi_2^\phi=\frac{1-x_{+}^{-\nu}}{\nu}\:Q_2\,.
\end{split}
\end{equation}
One can directly check that these quantities satisfy the first law of thermodynamics
\begin{equation}
d M_\phi\=T\,dS_\phi+\Phi_1\,dq_1^\phi+\Phi_2\,dq_2^\phi\;.
\end{equation}
The fundamental point for our discussion is that the boundary conditions \eqref{eq:bc} of this configuration allow for the possibility of having $\varphi_{0}=\varphi_{1}=0$. Indeed, the Reissner-Nordstr\"{o}m black hole satisfies these boundary conditions, therefore it is a  state in the same theory different than the hairy black hole.\par\smallskip
Let us study now the extremal limit of this hairy black hole.

\subsection{The supersymmetric and non-supersymmetric extremal hairy black hole}
In \cite{Anabalon:2020pez} we demonstrated that there exist electrically charged BPS black holes of finite horizon area for $\alpha^{2}=L^{-2}$ and, in the current parametrization, for the following values of the charge parameters:%
\footnote{%
as discussed in App.\ A of \cite{Gallerati:2021cty}, this configuration is 1/4 BPS
}
\begin{equation}
Q_1=-\,Q_2\;\frac{\numinus}{\nuplus}+\frac{k\,\mu}{\nuplus}\:,
\qquad\qquad
Q_2=\frac{\left(1+\nu+k\,\mu^2\right)\,\numinus}{2\,\mu}\:.
\end{equation}
In this case, the lapse function has a double zero as expected
\begin{equation}\label{k}
\begin{split}
&f(x)=\:\frac{x^{2-2\,\nu}}{4\,\nu^4}\,
    \Big((-1+x^\nu)^2\,k\,\mu^2+\big(2\,x^\nu(-1+\nu^{2})+x^{2\,\nu}(1-\nu)+\nu +1\big)\Big)^2\,,
\\[2ex]
&f\left(x_{+}\right)=0 \quad\Longrightarrow\quad\,
    x_{\pm}^{\nu}=\frac{1+k\,\mu^2-\nu\left(\nu\pm\sqrt{-1-2\,k\,\mu^2+\nu^2}\right)}{1+k\,\mu^2-\nu}\;,
\end{split}
\end{equation}
the BPS thermodynamical quantities of \cite{Anabalon:2020pez} being expressed in our parametrization as:
\begin{equation}
\begin{alignedat}{3}
q_1^{\phi}&=\frac{\sigma_{k}}{8\pi G}\:\frac{L}{\mu}\,
    \left(1+k\,\mu^2-\nu\right)\,\frac{\nuplus}{2\,\mu}\,,\qquad
&&\Phi_{1}^{\phi}=
    \frac{-1+x_{+}^\nu}{\nu}\left(1+k\,\mu^2-\nu\right)\,\frac{\nuplus}{2\,\mu}\,,
\\[\jot]
q_2^{\phi}&=\epsilon_2\,\frac{\sigma_{k}}{8\pi G}\:\frac{L}{\mu}\,
    \left(1+k\,\mu^2+\nu\right)\,\frac{\numinus}{2\,\mu}\,,\qquad\;\;
&&\Phi_{2}^{\phi}=\epsilon_2\,
    \frac{1-x_{+}^{-\nu}}{\nu}\left(1+k\,\mu^2+\nu\right)\,\frac{\numinus}{2\,\mu}\,,
\\[2\jot]
M^\phi&=\frac{\sigma_{k}}{8\pi G}\:L\,\frac{-1+\nu^{2}}{3\,\mu^{3}}\,,
\end{alignedat}
\label{eq:bps}
\end{equation}
where $\epsilon_{2}=\pm 1$. In particular, these quantities characterize a
supersymmetric black hole when $\epsilon_{2}=1$, and an extremal
non-supersymmetric hairy black hole when $\epsilon_{2}=-1$. It is possible to verify that $x_{\pm}(\nu)=x_{\pm}(-\nu)$, namely, $x_{\pm}$ is an even function of $\nu$. Hence, without loss of generality, it is possible to assume that $\nu\geq1$. The extremality condition implies now,
\begin{equation}
\delta M^{\phi}\=\Phi_{1}^{\phi}\:\delta q_{1}^{\phi}
    \+\Phi_{1}^{\phi}\:\delta q_{2}^{\phi}\:.
\end{equation}

\section{Canonical ensemble}
In order to compare the hairy black hole with the Reissner-Nordstr\"{o}m
solution in the canonical ensemble, we set their charges \eqref{eq:bps} and \eqref{eq:Mq1q2} at the same value:
\begin{equation}
q_{1}^{\phi}=q_{1}\,, \qquad\;\;  q_{2}^{\phi}=q_{2}\,.
\end{equation}
This yields the following relation between the parameters of the hairy
solution
\begin{equation}
\frac{q_{1}}{q_{2}}=\frac{q_{1}^{\phi}}{q_{2}^{\phi}}
\;\;\quad\Rightarrow\quad\;
\epsilon_{1}\,\frac{\nuplus}{\numinus}
    \=\epsilon_{2}\,\frac{\left(1+k\,\mu^2-\nu\right)}{\left(1+k\,\mu^2+\nu\right)}\:\frac{\nuplus}{\numinus}\,.
\label{relation}
\end{equation}
A detailed study shows that \eqref{relation} is a non-trivial result only
when \,$\epsilon_{1}=-\epsilon_{2}$\, and \,$k=-1$. In this case, we find that $\mu^{2}=1$. It follows from \eqref{k} that
\begin{equation}
x_{\pm}^{\nu}\=\nu\pm\sqrt{1+\nu^{2}}\:,
\end{equation}
and consequently
\begin{equation}
q_{1}^{\phi}\=q_{1}\=
    -\,\frac{\sigma_{-1}}{8\pi G}\;\nuplus\,L\:\frac{\nu}{2}\:,
\end{equation}
which implies that $q=-\frac{\nu}{2}$. This can be used to compute the
difference between the free energies of the Reissner-Nordstr\"{o}m black
hole and the hairy black hole as a function of $\nu$:
\begin{equation}
\Delta(\nu)\,
    =\left(F^{\phi}-F^\textsc{rn}\right)\,\frac{8\pi G}{L\,\sigma_{-1}}\:.
\end{equation}
We obtain that
\begin{equation}
\begin{alignedat}{3}
\nu=1            & \quad\Rightarrow\quad\; && \Delta(\nu)=0 \:, \\[2\jot]
1<\nu<\nu^{\ast} & \quad\Rightarrow\quad\; && \Delta(\nu)<0 \:, \\[2\jot]
\nu=\nu^{\ast}   & \quad\Rightarrow\quad\; && \Delta(\nu)=0 \:, \\[2\jot]
\nu>\nu^{\ast}   & \quad\Rightarrow\quad\; && \Delta(\nu)>0 \:,
\end{alignedat}
\end{equation}
with $\nu^{\ast}=\sqrt{1+\frac{2}{\sqrt{3}}}$\:. Hence, there is a set of
theories where hyperbolic supersymmetric black holes are unstable in the
canonical ensemble, namely, whenever $\nu>\nu^{\ast}$\, (see also \eqref{eq:SO8breaking}).

\section{Discussion}
We conclude that there exists a set of supersymmetric black holes which can be unstable in the canonical ensemble. For this to happen, it is crucial that these black holes feature a horizon with locally hyperbolic geometry ($k=-1$). We would also like to remark that we have carried out the same analysis in the grand canonical ensemble and we have obtained that, in this case, all the supersymmetric black holes are stable, independently of the topology. The same happens for the planar ($k=0$) and spherical ($k=1$) topologies in the canonical ensemble.\par
Since the geometry of the horizon is locally hyperbolic, we shall further discuss the global features of this geometry, in order to provide a sharper understanding of the phenomenon. To this end, we shall divide the discussion in two cases: the compact horizon and the non-compact one.

\subsection{Non-compact horizon}
In this case, the geometries are actually asymptotic AdS. Therefore, the geometry of the Reissner-Nordstr\"{o}m black hole has the same asymptotic Killing spinors as the AdS${}_4$ itself. When there is no running scalar field, the only supersymmetric solution within the class of metrics considered here is AdS${}_4$ \cite{Caldarelli:1998hg}. The hyperbolic Reissner-Nordstr\"{o}m black hole with non-compact horizon can be then interpreted as an excited state over this SUSY ground state, the latter being reached in the limit of vanishing mass and charge.\par
The hairy supersymmetric black hole has a different set of Killing spinors that are not those of globally AdS${}_4$, not even asymptotically. This means that the supersymmetry algebra has a different realization for the hairy black holes and the BPS bound they satisfy allows the mass to be larger than zero. The Killing spinors of the hairy black holes are a function of the radial coordinate only and, therefore, the algebra has a realization that is similar to mAdS \cite{Hristov:2011ye, Hristov:2011qr}. Hence, it is not unusual that the BPS solution can have a larger energy compared with a non-BPS one, each geometry being actually connected with a different representation of the supersymmetry algebra.

\subsection{Compact horizon}
In this case, the geometries are asymptotically locally AdS. The locally hyperbolic horizon is compactified to a surface of genus $g$ with volume $\sigma_{-1}$. This asymptotic geometry has no Killing spinors at all, so the electrically charged, hyperbolic Reissner-Nordstr\"{o}m black hole with non-compact horizon is not connected with any representation of a supersymmetry algebra. The hairy supersymmetric black hole still has Killing spinors when the horizon is compact. Therefore, the configuration that saturates the BPS bound can be more energetic than other geometries, with same charges and boundary conditions, which can not be seen as excited states over a BPS ground state.\par\bigskip
The comparison between the two class of solutions (hairy BPS and extremal RN) in the canonical ensemble pointed out a thermodynamic instability of the hairy supersymmetric black hole for a certain range of parameters. We still do not know if this would also imply a specific dynamical instability of the solution, but we expect the more favourable Reissner-Nordström configuration as the final state of a thermodynamic phase transition for $\nu>\nu^{\ast}$.
We are not aware of any observation in the literature in which this instability of the supersymmetric states has been reported before. We also believe that the analysis performed here claims for a deeper understanding of SUSY algebra in the presence of electromagnetic gaugings, following the lines of \cite{Hristov:2011ye, Hristov:2011qr}. It also seems that the electromagnetic gauging plays a role similar to electric and magnetic charges, yielding a new representation of the SUSY algebra. The BPS state turns out to be solitonic in this regard. We leave the clarification of these issues to a forthcoming work.

\section*{\protect\normalsize Acknowledgments}
\vspace{-5pt}
We would like to thank the support of Proyecto de cooperación internacional 2019/13231-7 FAPESP/ANID. The research of AA is supported in part by the Fondecyt Grants 1210635 and 1181047. The research of DA is supported in part by the Fondecyt Grant 1200986.



\hypersetup{linkcolor=blue} \phantomsection 
\addcontentsline{toc}{section}{References}
\bibliographystyle{mybibstyle}
\bibliography{bibliografia}

\end{document}